\documentclass[cameraready]{Interspeech}
\usepackage{subcaption}
\usepackage{booktabs}
\usepackage{multirow} % for multirow cells
\usepackage{adjustbox} % for resizing
\usepackage{caption}
\usepackage{tabularx}
\usepackage{cite}
\usepackage{xcolor}
\usepackage[table]{xcolor}

\title{Hearing the Order: Investigating Position Bias \\ in Large Audio-Language Models}

\author[equalcontribution]{Yu-Xiang}{Lin$^1$}
\author[equalcontribution]{Chen-An}{Li$^1$}
\author{Sheng-Lun}{Wei$^1$}
\author{Po-Chun}{Chen$^1$}
\author{Hsin-Hsi}{Chen$^{1, 2}$}
\author{Hung-yi}{Lee$^{1, 3}$}
\address{
  \textsuperscript{1}National Taiwan University \\
  \textsuperscript{2}NTU AI Research Center (AINTU) \\
  \textsuperscript{3}NTU Artificial Intelligence Center of Research Excellence (NTU AI-CoRE)
}

\email{f13942075@ntu.edu.tw, f13942069@ntu.edu.tw, hungyilee@ntu.edu.tw}

\keywords{Large Audio-Language Model, Position Bias}

\usepackage{comment}

\begin{document}

\maketitle

\begin{abstract}
    Large audio-language models (LALMs) are often used in tasks that involve reasoning over ordered options. An open question is whether their predictions are influenced by the order of answer choices, which would indicate a form of position bias and undermine their reliability. In this paper, we identify and analyze this problem in LALMs. We demonstrate that no model is immune to this bias through extensive experiments on six LALMs across three widely used benchmarks and their spoken counterparts. Shuffling the order of answer options can cause performance fluctuations of up to 24\% and even change model rankings, raising concerns about the reliability of current evaluation practices. We also study permutation-based strategies and show that they can mitigate bias in most cases. Our work represents the first systematic investigation of this issue in LALMs, and we hope it raises awareness and motivates further research in this direction.
\end{abstract}

\section{Introduction}
Large audio-language models (LALMs)~\cite{openai2024gpt4ocard, comanici2025gemini, abouelenin2025phi, xu2025qwen2, liu2025voxtral, lu2025desta2} have advanced rapidly in recent years, demonstrating strong audio understanding and reasoning capabilities~\cite{huang2024dynamic, sakshimmau, ma2025mmar, kumar2025mmau, yang2025towards, yang25g_interspeech}. 
Among the various evaluation formats, multiple-choice question (MCQ) benchmarks have become particularly popular, as they provide standardized options and allow precise matching of answers, facilitating fair comparisons across models. 
However, MCQs also conceal an important concern: model decisions may be influenced by the order in which options are presented rather than solely by their semantic content, a phenomenon known as position bias. 
While this issue has been surveyed in text-only language models~\cite{izacard2023atlas, zhang2024found, wei-etal-2024-unveiling, liusie-etal-2024-teacher, xu-etal-2024-context, pezeshkpour2024large, li-gao-2025-anchored, choi-etal-2025-mitigating} and vision-language models~\cite{tian2025identifying, loginova-etal-2025-addressing}, where studies show systematic preferences for certain option positions. 
However, its impact on LALMs remain largely unexplored.

In this work, we comprehensively study six LALMs across three representative MCQ benchmarks and their spoken counterparts. 
Our experiments show that position bias is indeed widespread. 
All models we tested are affected, and none are free from this flaw. 
Our findings suggest that common evaluation methods for LALMs may introduce unnecessary bias and yield results that do not fully reflect the models' true capabilities. 
Simply relying on conventional evaluation could give a misleading picture of model performance. Our results show that permutation-based strategies, although computationally intensive, can effectively reduce position bias and provide more reliable evaluations of LALMs. 
To the best of our knowledge, we are the first paper to investigate this kind of problem in LALMs. 
We hope this work raises awareness of the problem and motivates further research into the development of specialized evaluation frameworks and mitigation methods.

\section{Related Work}

\subsection{Multiple-Choice Question Benchmark}
Multiple-choice questions (MCQs) have long been a common and effective approach for evaluating large language models (LLMs). 
This method transforms open-ended generative tasks into classification problems, reducing the reliance on costly human grading and ensuring objectivity and comparability across models. 
Representative benchmarks for LLMs include ARC-Challenge~\cite{clark2018think} and MMLU~\cite{hendryckstest2021}, which have become standard tools for measuring reasoning and knowledge coverage.

Also, a growing body of benchmarks has been proposed to evaluate LALMs~\cite{yang2025towards}. 
For example, Dynamic-SUPERB~\cite{huang2024dynamic} and AIR-Bench~\cite{yang2024air} assess LALMs' ability to comprehend diverse audio signals. 
MMAU~\cite{sakshimmau}  introduces human-annotated questions that demand expert-level knowledge and multi-step reasoning. 
Building on this, MMAR~\cite{ma2025mmar} expands the evaluation to a broader range of real-world auditory scenarios. 
Similarly, SAKURA~\cite{yang25g_interspeech} focuses on multi-hop reasoning, requiring models to recall and connect multiple facts across different audio contexts. More recently, SpeechR~\cite{yang2025speechr} and MMAU-Pro~\cite{kumar2025mmau} have been proposed to capture even more challenging aspects of audio-language understanding.

Although these benchmarks differ in design goals and targeted skills, they are all formulated as MCQs. This consistency highlights the strengths of MCQs for standardized evaluation. 
However, they also have limitations and the risk of position bias. 
Thus, while MCQ-based benchmarks have been instrumental in driving progress, there remains an urgent need for more robust evaluation methodologies that can better capture the full spectrum of LALM capabilities.

\subsection{Position Bias in LLMs}
Early research on LLMs has primarily focused on position bias, showing that where a passage appears within a long context can substantially alter model outputs~\cite{kazemnejad2023impact, liu-etal-2024-lost, he-etal-2024-never, zhang2024found}. 
Such positional effects raise concerns about the reliability of LLMs as evaluators~\cite{wang-etal-2024-large-language-models-fair, li2024split, zhu2025judgelm, verga2024replacing}. 
In in-context learning, the ordering of exemplars further demonstrates that model behavior can be sensitive to positional effects~\cite{lu-etal-2022-fantastically, zhang-etal-2024-batch, xu-etal-2024-context}.

More recent work has expanded this view from positional effects to the broader notion of position bias. 
In multiple-choice settings, for instance, both the order of candidate options and the identifiers assigned can significantly change accuracy~\cite{zheng2024large, pezeshkpour2024large, wang-etal-2024-answer-c, reif-schwartz-2024-beyond, li-gao-2025-anchored, xue-etal-2024-strengthened, wei-etal-2024-unveiling}. 
To counter these biases, cyclic and full-permutation methods have been proposed~\cite{izacard2023atlas, zheng2024large, liusie-etal-2024-teacher}. 
These approaches, akin to self-consistency~\cite{wangself}, average predictions across permutations. 
Although they introduce additional test-time compute, they consistently reduce position bias and improve evaluation reliability.

Beyond text-only LLMs, emerging evidence suggests that position bias also arises in multimodal models. 
For example, research on Large Vision-Language Models and Video Language Models shows that positional effects influence predictions~\cite{tian2025identifying, loginova-etal-2025-addressing}. 
However, to our knowledge, no systematic study of position bias in LALMs has been conducted. 
Our work addresses this gap by providing the first comprehensive analysis.

\section{Identifying Position Bias in LALMs}
\subsection{Experiment Setup}
\subsubsection{Benchmarks} 
We conduct the experiments on MMAU~\cite{sakshimmau}, MMAR~\cite{ma2025mmar}, and MMLU~\cite{hendryckstest2021}, as they are widely recognized and representative benchmarks for evaluating LALMs. For MMAU, we use the test-mini subset, since only its answers are publicly available, which enables further analysis. 
To study modality effects, we construct SPEECH-MMAU, SPEECH-MMAR, and SPEECH-MMLU by converting textual questions and their options into speech using GPT-4o mini TTS\footnote{\url{https://platform.openai.com/docs/models/gpt-4o-mini-tts}}, which is based on GPT-4o mini~\cite{openai2024gpt4ocard} and produces natural-sounding speech and handles complex pronunciation well~\cite{manku2025emergenttts}. 

Since the spoken conversion substantially increases sequence length, some audio samples become extremely long. 
To maintain tractability under our computational constraints, we filter out those exceeding 180 seconds. 
Also, we retain  samples with exactly four answer options to ensure comparability. 
After these filtering steps, the statistics of the resulting test sets are reported in Table~\ref{tab:answer-distribution}.
 
Beyond dataset preparation, we also aim to analyze potential position bias. 
To this end, we systematically reassign the correct answer to each of the four option positions A, B, C, and D, while randomly shuffling the remaining options. 
Under each setting, the correct answer is fixed at the designated position. 
This procedure ensures that across settings the correct answer is fully covered at every position, allowing us to examine how model accuracy changes with positional effects under directly comparable conditions.

\begin{table}[h]
    \centering
    \caption{Number of samples per answer position (A–D) in each dataset, with proportions in parentheses.}
    \label{tab:answer-distribution}
    \resizebox{\linewidth}{!}{
    \begin{tabular}{lccccc}
        \toprule
         & A & B & C & D & Total \\
        \midrule
        MMAU   & 357 (38.3\%) & 257 (27.5\%) & 201 (21.5\%) & 118 (12.6\%) & 933 \\
        MMAR   & 187 (22.9\%) & 209 (25.6\%) & 208 (25.5\%) & 211 (25.9\%) & 815 \\
        MMLU   & 3213 (22.9\%) & 3458 (24.7\%) & 3577 (25.5\%) & 3771 (26.9\%) & 14019 \\
        \bottomrule
    \end{tabular}
    }
    %\vspace{-10pt}
\end{table}

\subsubsection{Models} 
To benchmark model performance under these conditions, we adopt six state-of-the-art LALMs capable of handling long audio:  Gemini-2.0-Flash~\cite{comanici2025gemini},  Phi-4-Multimodal~\cite{abouelenin2025phi},  Qwen2.5-Omni-3B~\cite{xu2025qwen2},  Qwen2.5-Omni-7B~\cite{xu2025qwen2},  Voxtral-Mini-3B~\cite{liu2025voxtral}, and  Voxtral-Small-24B~\cite{liu2025voxtral}. 
These models are chosen to cover different architectures and a range of model sizes, enabling ablation studies on architectural diversity and scaling. 
We follow the OpenAI simple-eval protocol\footnote{\url{https://github.com/openai/simple-evals}} to instruct the models, fixing the temperature at 0 for reproducibility and setting the maximum output sequence length to 1024 tokens.

\subsubsection{Metrics}
For evaluation, we report several types of metrics. First, we use accuracy, which is defined as the proportion of correctly answered samples relative to the total number of samples in a dataset. Second, we report $\Delta$ accuracy, which measures the difference in accuracy between the original dataset and the reassigned-answer setting, thereby quantifying robustness to positional shuffling. Finally, we adopt two widely used measures to analyze position bias, Relative Standard Deviation (RSD)~\cite{reif2024beyond} and Choice Kullback-Leibler Divergence (CKLD)~\cite{choi-etal-2025-mitigating}. These metrics are formally defined as: 
\begin{equation}
\mathrm{RSD} = \frac{\sqrt{\tfrac{1}{k}\sum_{i=1}^k (s_i - \bar{s})^2}}{\bar{s}}, 
\quad
\mathrm{CKLD} = \sum_{i=1}^{k} p_i \log \frac{p_i}{q_i}, 
\end{equation}
where $k$ denotes the number of choices, $s_i$ is the accuracy of the $i$-th choice, and $\bar{s}$ is the mean accuracy across choices. In the CKLD metric, $p_i$ represents the proportion of predictions for the $i$-th choice, while $q_i$ denotes the proportion of the ground-truth label. We use these two metrics for a comprehensive evaluation, following~\cite{choi-etal-2025-mitigating}.

\begin{figure*}[!htbp]
    \centering
    \begin{subfigure}{0.32\linewidth}
        \centering
        \includegraphics[height=120pt,width=\linewidth]{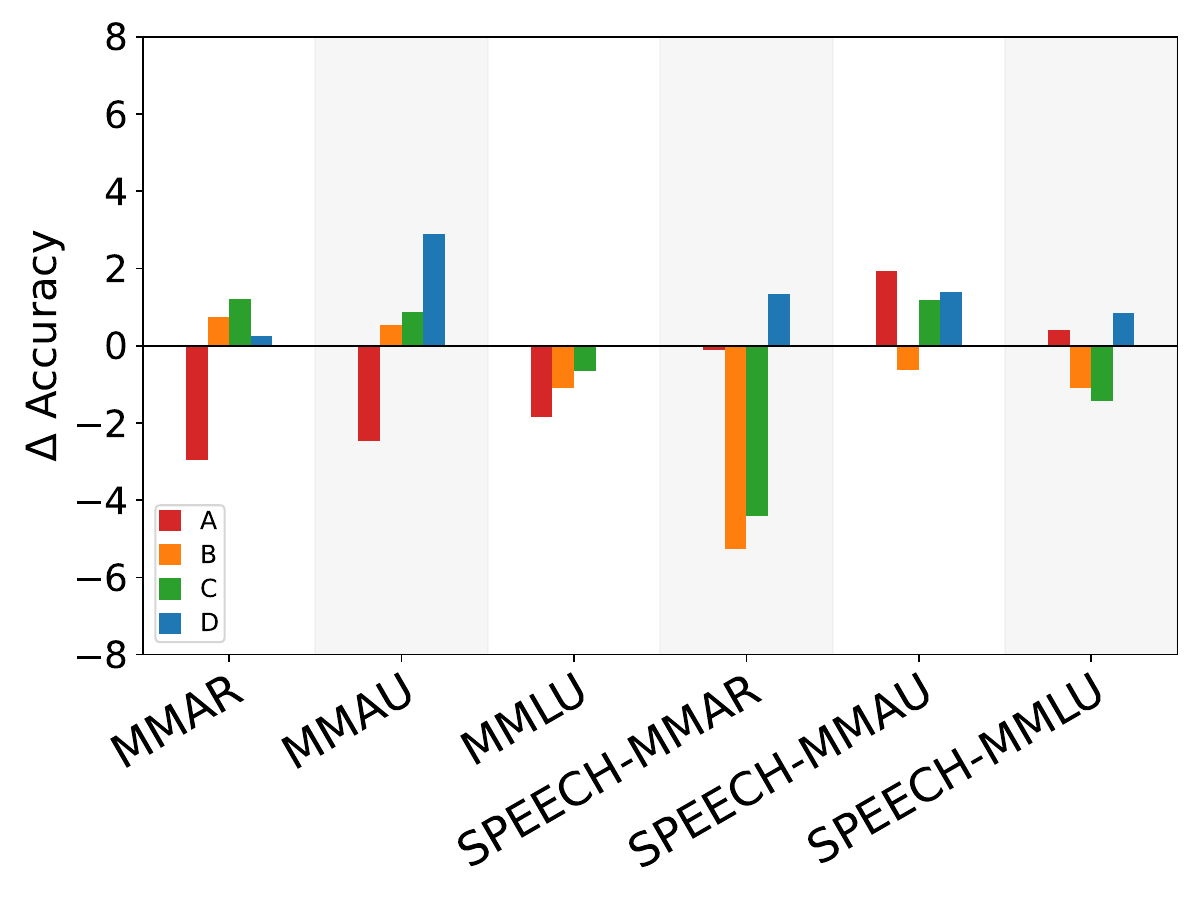}
        \subcaption{Gemini-2.0-Flash}
    \end{subfigure}
    \begin{subfigure}{0.32\linewidth}
        \centering
        \includegraphics[height=120pt,width=\linewidth]{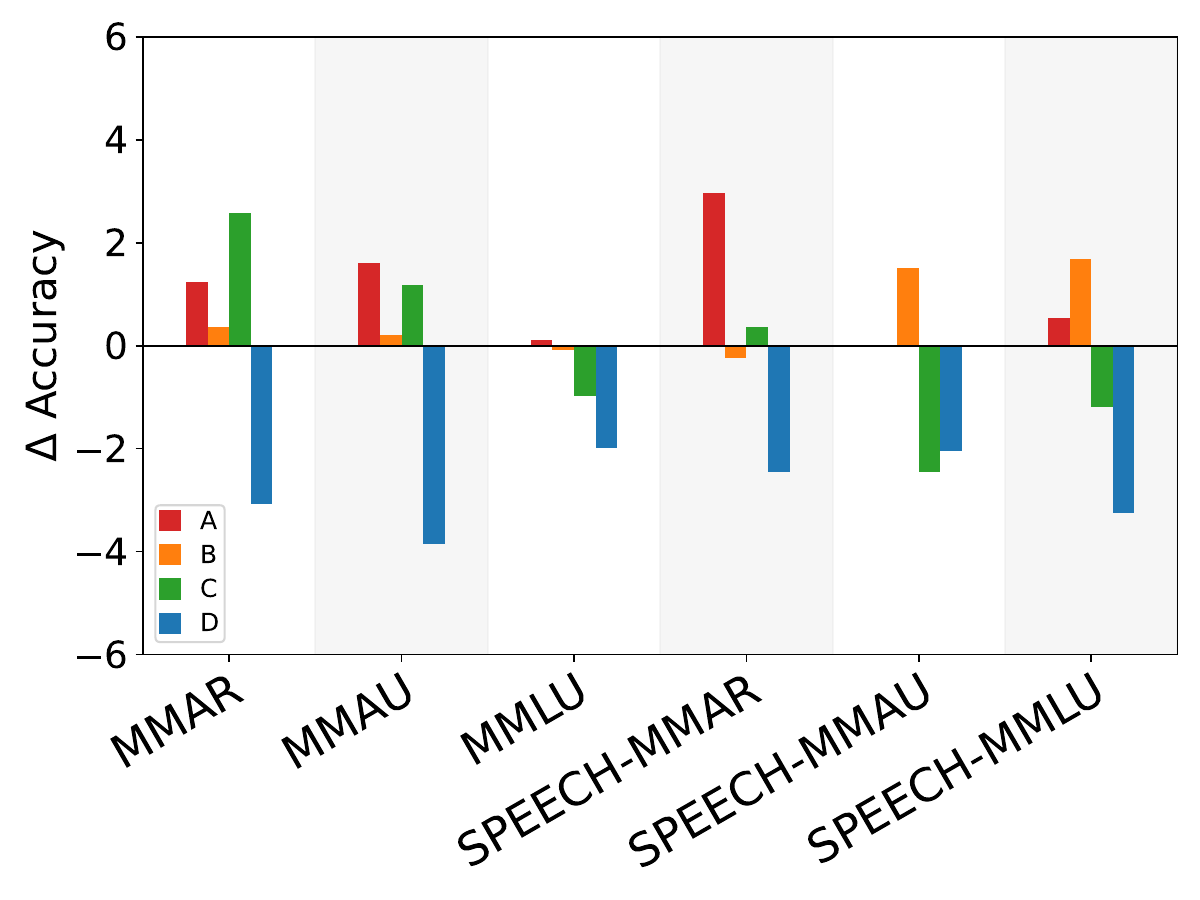}
        \subcaption{Qwen2.5-Omni-3B}
    \end{subfigure}
    \begin{subfigure}{0.32\linewidth}
        \centering
        \includegraphics[height=120pt,width=\linewidth]{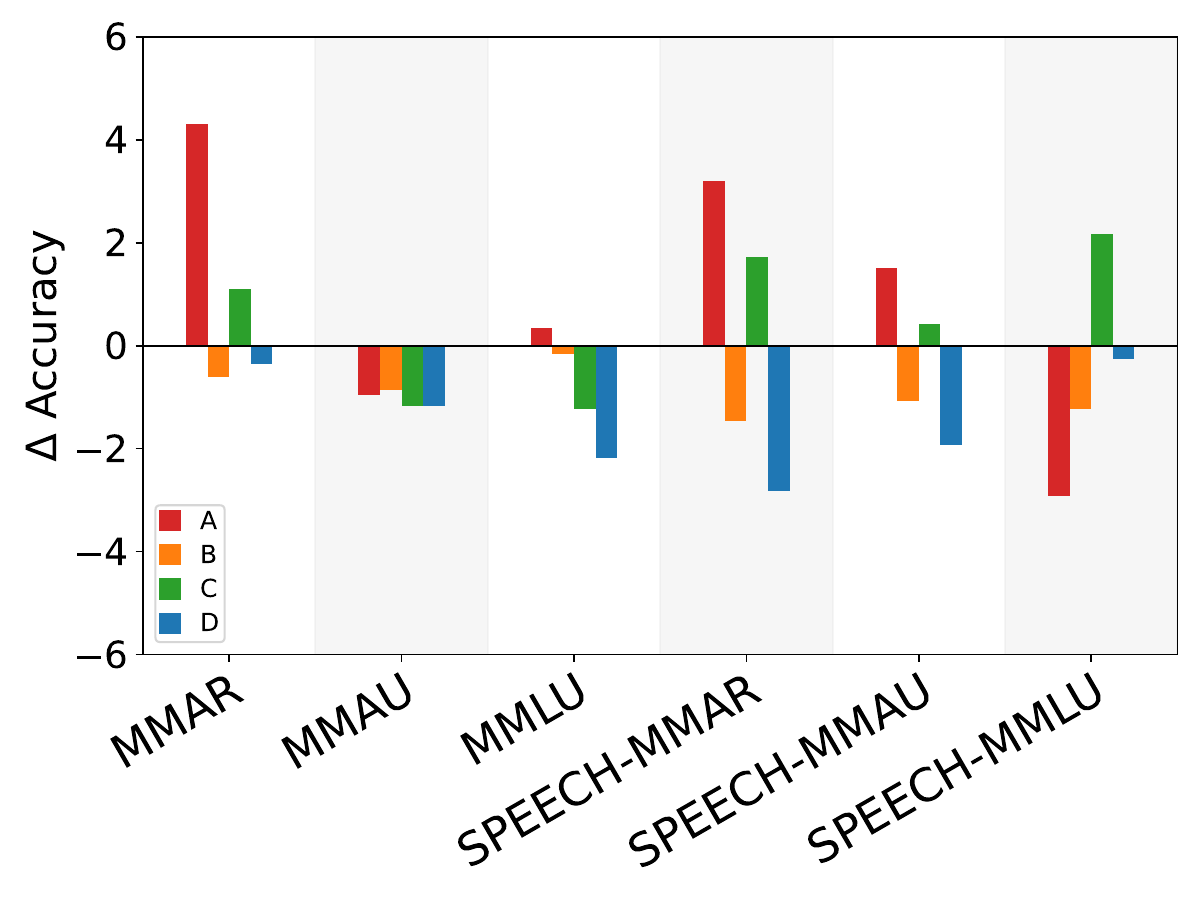}
        \subcaption{Qwen2.5-Omni-7B}
    \end{subfigure}
    \hfill    
    \begin{subfigure}{0.32\linewidth}
        \centering        \includegraphics[height=120pt,width=\linewidth]{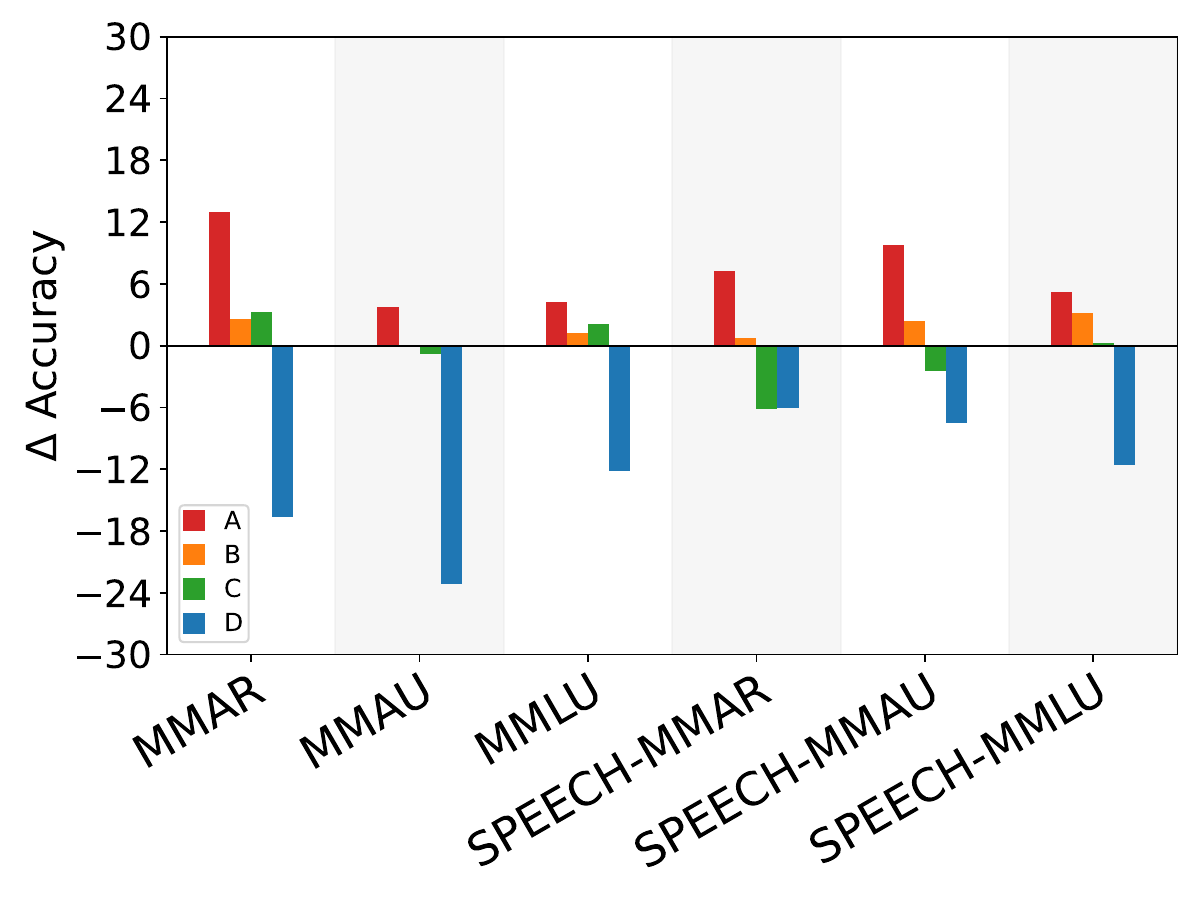}
        \subcaption{Phi-4-Multimodal}
    \end{subfigure}
    \vspace{0.5em}  
    \begin{subfigure}{0.32\linewidth}
        \centering
        \includegraphics[height=120pt,width=\linewidth]{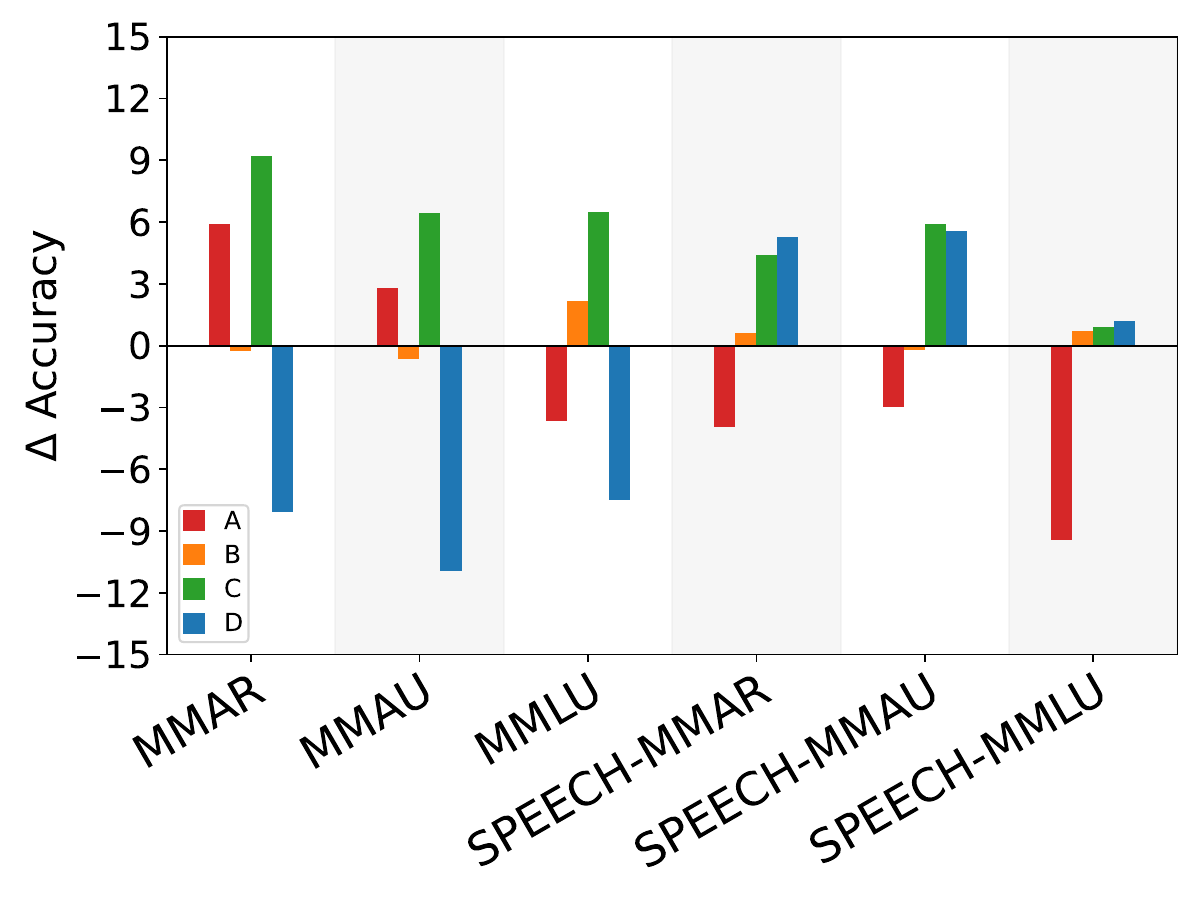}
        \subcaption{Voxtral-Mini-3B}
    \end{subfigure}
    \begin{subfigure}{0.32\linewidth}
        \centering
        \includegraphics[height=120pt,width=\linewidth]{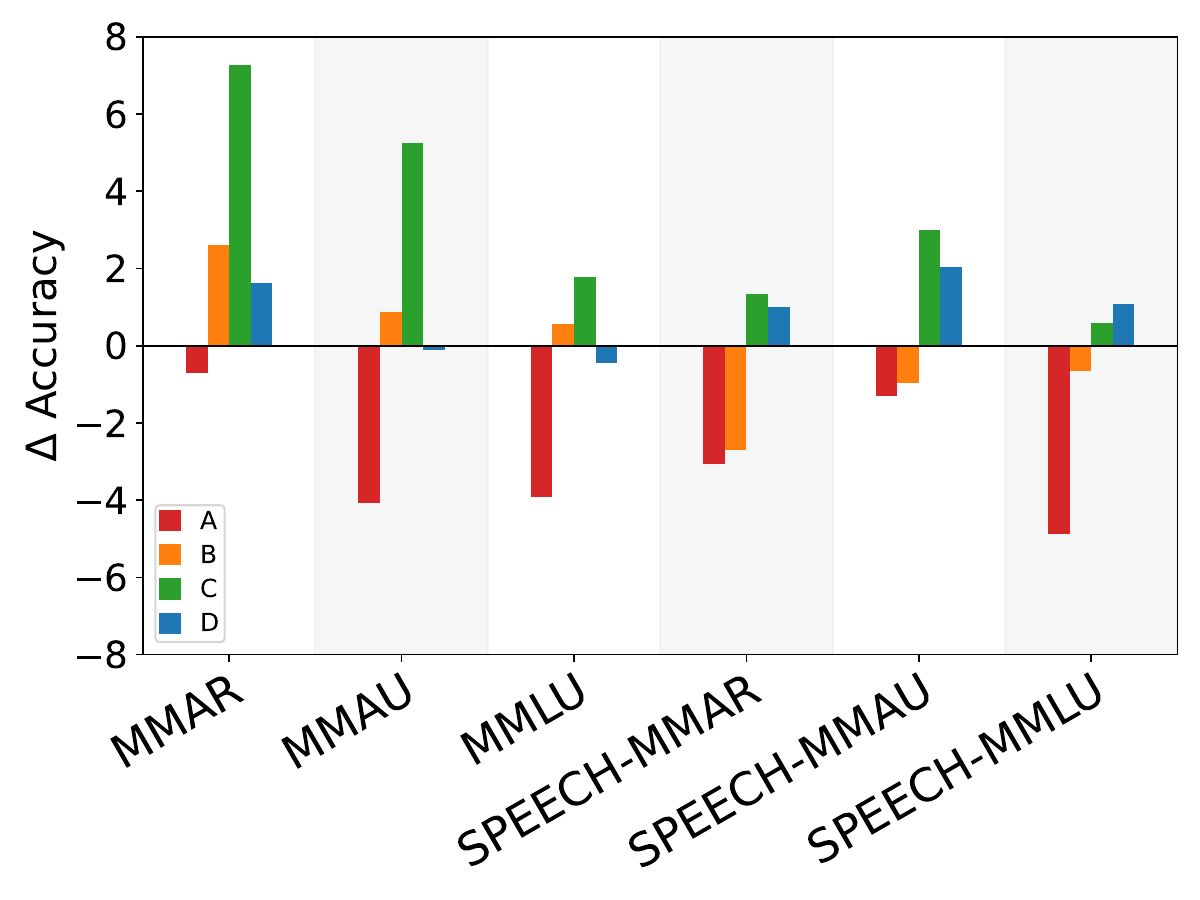}
        \subcaption{Voxtral-Small-24B}
    \end{subfigure}
    % \vspace{-10pt}
    \caption{Performance difference ($\Delta$ Accuracy) across different datasets when the correct answer is systematically reassigned to a fixed option position (A, B, C, or D). This highlights how model accuracy changes depending on the location of the correct answer choice.}
    % \vspace{-10pt}
    \label{fig:fluctuation}
\end{figure*}

\subsection{Results}
In Figure~\ref{fig:fluctuation}, we report $\Delta$ accuracy, defined as the difference in performance between the original dataset and the reassigned-answer setting to evaluate not only which option a model tends to prefer, but also the magnitude of that preference. 
Our experiments span six datasets, MMAU, SPEECH-MMAU, MMLU, SPEECH-MMLU, MMAR, and SPEECH-MMAR, providing a comprehensive analysis of position bias.

From Figure~\ref{fig:fluctuation}, it is evident that every evaluated model demonstrates systematic fluctuations in accuracy when the correct answer is reassigned to a fixed option position, revealing that option position bias is a pervasive phenomenon. 
Across all datasets, no model is exempt from this effect, underscoring its severity as a fundamental challenge. While some models appear relatively less sensitive, the magnitude of the fluctuations makes clear that robustness remains unsatisfactory overall. 
For instance,  Gemini-2.0-Flash,  Qwen2.5-Omni-3B, and  Qwen2.5-Omni-7B exhibit accuracy variations of approximately 5\%. 
In contrast, models such as Voxtral-Mini-3B, Voxtral-Small-24B, and Phi-4-Multimodal exhibit much more pronounced bias. 
In particular, Phi-4-Multimodal reaches a maximum fluctuation of nearly 24\%, illustrating how drastically the arbitrary placement of the correct option can sway model outputs.

A closer inspection shows that each model exhibits clear and consistent bias. 
For example,  Phi-4-Multimodal often favors answers in position A and strongly avoids position D, leading to large fluctuations. 
Voxtral-Mini-3B tends to avoid D in text datasets but prefers D in speech datasets. 
Despite similar training data and design,  Voxtral-Small-24B behaves differently, consistently disfavoring A across text and speech. 
In contrast,  Qwen2.5-Omni-3B and  Qwen2.5-Omni-7B show similar patterns: both avoid D and show a relative preference for A. 
Finally,  Gemini-2.0-Flash shows the opposite trend, generally preferring D.

In sum, these findings highlight that although the severity and direction of position bias vary across models, the problem itself is universal. 
Regardless of architecture, scale, or training corpus, all models investigated here remain  vulnerable to systematic fluctuations caused by option placement. 
This demonstrates that position bias is not only an artifact of individual models but a structural weakness across current LALMs.

\section{In-depth Analysis}
\subsection{Effect of Identifiers}
We investigate whether model behavior is more influenced by the order of the correct answer or by the identifiers assigned to each option. 
To test this, we evaluate models with and without standard identifiers A, B, C, and D, as shown in Table~\ref{tab:without_id_eval}.

Across MMAU, identifiers tend to improve model accuracy in most cases, though they do not reliably reduce bias. On SPEECH-MMAU, identifiers similarly enhance stability in many instances, but this benefit does not translate into a consistent reduction of fluctuations. 
While performance gains are often observed, the effect on bias remains negligible.

Overall, these findings show that identifiers achieve better accuracy than without them, but they do not mitigate position bias. 
Their impact varies across model families and modalities, reflecting the complex interaction between answer order and identifiers.

\begin{table}[hbtp]
    \centering
    \caption{Model performance with and without option identifiers on MMAU and SPEECH-MMAU.}
    \resizebox{0.9\linewidth}{!}{
        \begin{tabular}{lcccccccc}
            \toprule
            \multirow{2}{*}{Model} & 
            \multicolumn{2}{c}{MMAU} & 
            \multicolumn{2}{c}{SPEECH-MMAU} \\
            \cmidrule(lr){2-3} \cmidrule(lr){4-5}
             & CKLD$\downarrow$ & Acc$\uparrow$ & CKLD$\downarrow$ & Acc$\uparrow$ \\
            \midrule
            Phi-4-multimodal & 0.0171 & 65.27 &  0.0080 &  53.06 \\
            \quad - Without ID & 0.0215 & 58.19 &  0.0727 &  39.42 \\
            \midrule
            Qwen2.5-Omni-7B & 0.0017 & 73.96 &  0.0040 &  71.92 \\
            \quad - Without ID & 0.0014 & 69.02 &  0.0045 &  58.20 \\
            \midrule
            Qwen2.5-Omni-3B & 0.0031 & 70.84 &  0.0015 &  68.17 \\
            \quad - Without ID & 0.0019 & 60.77 &  0.0104 &  51.55 \\
            \midrule
            Voxtral-Mini-3B & 0.0055 & 54.98 &  0.0251 &  54.23 \\
            \quad - Without ID & 0.0130 & 54.87 &  0.0178 &  44.91 \\
            \bottomrule
        \end{tabular}
    }
    \label{tab:without_id_eval}
    \vspace{-10pt}
\end{table}

\subsection{Comparing Position Bias in LALMs and LLMs}
\label{sec:lalms_llms}
\begin{figure}[htbp]
    \centering
    \begin{subfigure}{0.49\linewidth}
        \centering
        \includegraphics[width=\linewidth]{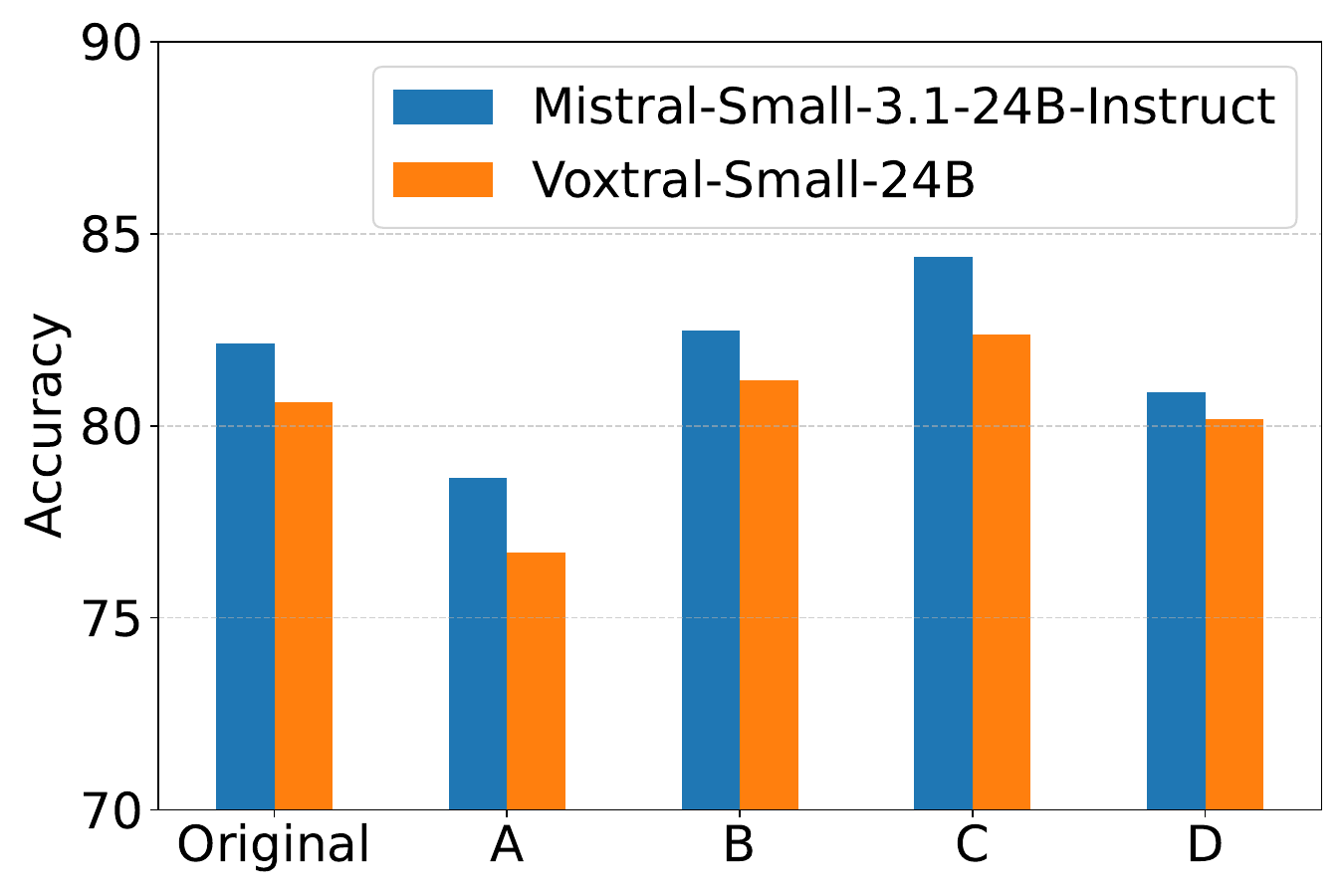}
    \end{subfigure}
    \begin{subfigure}{0.49\linewidth}
        \centering
        \includegraphics[width=\linewidth]{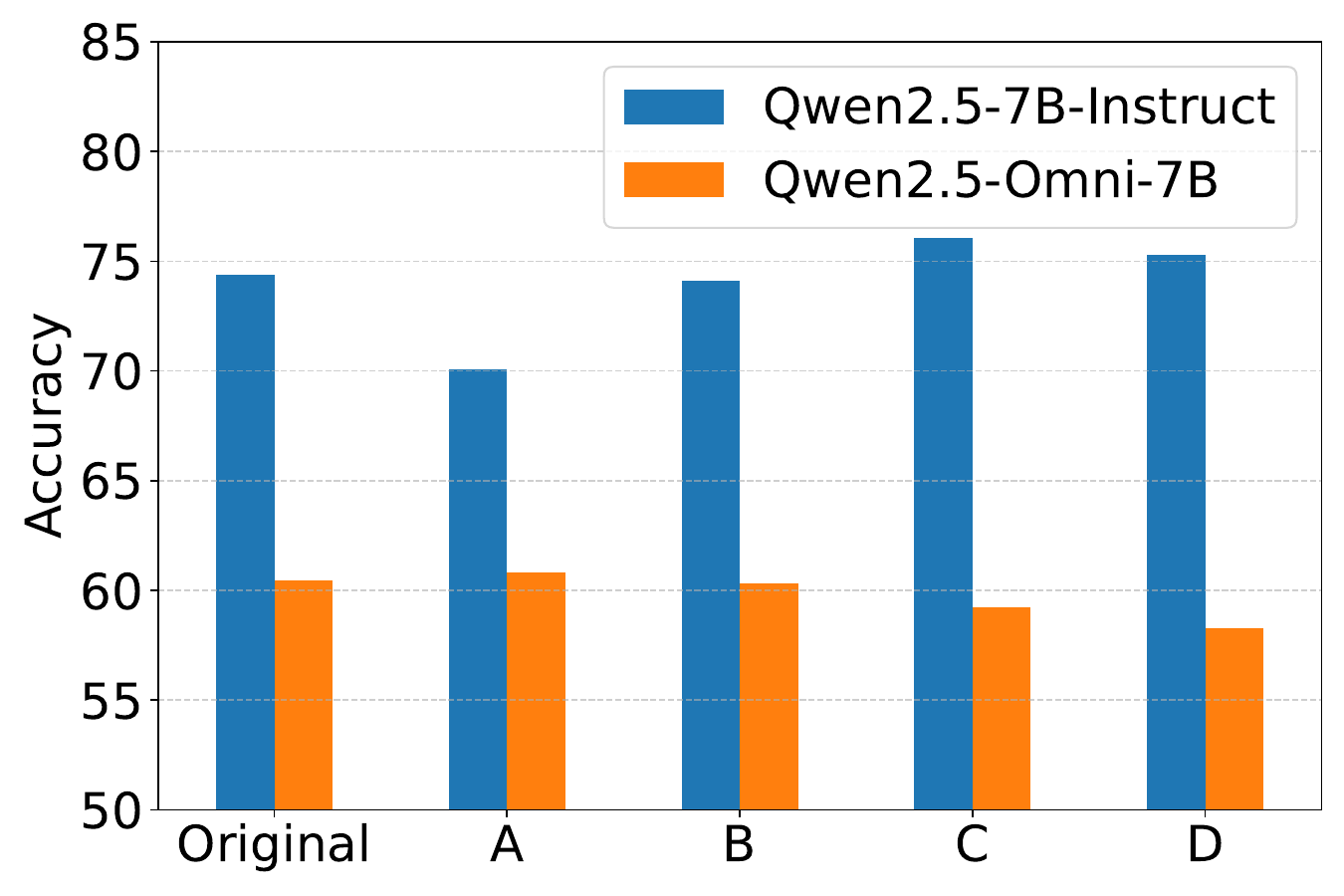}
    \end{subfigure}
    %\vspace{-10pt}
    \caption{Comparison of LALMs and their text-only LLM counterparts on MMLU}
    \label{fig:text_speech}
    \vspace{-15pt}
\end{figure}

This section compares LALMs with their text-only LLM counterparts to examine whether position bias is inherited from the base model or altered after audio-language instruction tuning. 
Figure~\ref{fig:text_speech} presents two comparisons on MMLU: Voxtral-Small-24B vs. Mistral-Small-3.1-24B-Instruct, and Qwen2.5-Omni-7B vs. Qwen2.5-7B-Instruct. 
We evaluate these models on MMLU and the reassigned-answer variants to examine accuracy and bias trends.

For Voxtral and Mistral, the trends in position bias are broadly similar, suggesting that the bias is mainly inherited from the text-only model. 
In contrast, the comparison of the  Qwen series shows a clear divergence, indicating that position bias is not always directly carried over. 
These results show that while some LALMs retain the position bias of their text-only counterparts, others exhibit distinct behaviors after fine-tuning.

\begin{table*}[htbp]
\centering
\caption{Permutation results across four datasets. Bold numbers indicate the best metric among the three permutations for each model-dataset pair. Cells with green background indicate RSD/CKLD improved over or equal to the original setting. Cells with purple background indicate Accuracy improved over the original setting.}
\vspace{-5pt}
\begin{adjustbox}{max width=\textwidth}
\begin{tabular}{ll*{4}{ccc}}
\toprule
\multirow{2}{*}{Model} & \multirow{2}{*}{Permutation} 
    & \multicolumn{3}{c}{MMAU} 
    & \multicolumn{3}{c}{SPEECH-MMAU} 
    & \multicolumn{3}{c}{MMAR} 
    & \multicolumn{3}{c}{SPEECH-MMAR} \\
\cmidrule(lr){3-5} \cmidrule(lr){6-8} \cmidrule(lr){9-11} \cmidrule(lr){12-14}
 & & RSD$\downarrow$ & CKLD$\downarrow$ & Acc$\uparrow$ & RSD$\downarrow$ & CKLD$\downarrow$ & Acc$\uparrow$ & RSD$\downarrow$ & CKLD$\downarrow$ & Acc$\uparrow$ & RSD$\downarrow$ & CKLD$\downarrow$ & Acc$\uparrow$ \\
\midrule
\multirow{3}{*}{Phi-4-multimodal}
 & Original   & 0.246 & 0.017 & 65.27 & 0.273 & \textbf{0.008} & 53.06 & 0.247 & 0.097 & 43.19 & 0.174 & 0.018 & 41.47 \\
 & Cyclic     & \cellcolor{green!20}{0.117} & \cellcolor{green!20}{\textbf{0.001}} & \cellcolor{blue!15}{69.02} & \cellcolor{green!20}{0.164} & \cellcolor{green!20}{\textbf{0.008}} & \cellcolor{blue!15}{54.34} & \cellcolor{green!20}{0.090} & \cellcolor{green!20}{0.014} & \cellcolor{blue!15}{48.59} & \cellcolor{green!20}{0.136} & \cellcolor{green!20}{0.010} & \cellcolor{blue!15}{44.17} \\
 & Full       & \cellcolor{green!20}{\textbf{0.104}} & \cellcolor{green!20}{0.002} & \cellcolor{blue!15}{\textbf{69.24}} & \cellcolor{green!20}{\textbf{0.110}} & 0.010 & \cellcolor{blue!15}{\textbf{58.31}} & \cellcolor{green!20}{\textbf{0.053}} & \cellcolor{green!20}{\textbf{0.007}} & \cellcolor{blue!15}{\textbf{51.53}} & \cellcolor{green!20}{\textbf{0.077}} & \cellcolor{green!20}{\textbf{0.001}} & \cellcolor{blue!15}{\textbf{46.50}} \\
\midrule
\multirow{3}{*}{Qwen2.5-Omni-7B}
 & Original   & 0.084 & 0.002 & 73.96 & 0.120 & 0.004 & 71.92 & 0.069 & 0.011 & 53.25 & 0.091 & 0.020 & 52.27 \\
 & Cyclic     & \cellcolor{green!20}{0.071} & \cellcolor{green!20}{\textbf{0.001}} & \cellcolor{blue!15}{75.99} & \cellcolor{green!20}{0.085} & \cellcolor{green!20}{0.003} & \cellcolor{blue!15}{74.28} & \cellcolor{green!20}{0.049} & \cellcolor{green!20}{0.009} & \cellcolor{blue!15}{55.21} & \cellcolor{green!20}{0.072} & \cellcolor{green!20}{0.010} & \cellcolor{blue!15}{54.48} \\
 & Full       & \cellcolor{green!20}{\textbf{0.056}} & \cellcolor{green!20}{0.002} & \cellcolor{blue!15}{\textbf{78.03}} & \cellcolor{green!20}{\textbf{0.062}} & \cellcolor{green!20}{\textbf{0.002}} & \cellcolor{blue!15}{\textbf{75.46}} & \cellcolor{green!20}{\textbf{0.023}} & \cellcolor{green!20}{\textbf{0.004}} & \cellcolor{blue!15}{\textbf{56.69}} & \cellcolor{green!20}{\textbf{0.037}} & \cellcolor{green!20}{\textbf{0.006}} & \cellcolor{blue!15}{\textbf{56.07}} \\
\midrule
\multirow{3}{*}{Qwen2.5-Omni-3B}
 & Original   & 0.101 & \textbf{0.003} & 70.84 & 0.111 & \textbf{0.002} & 68.17 & 0.061 & 0.006 & 51.53 & 0.136 & 0.017 & 51.54 \\
 & Cyclic     & \cellcolor{green!20}{0.045} & 0.004 & \cellcolor{blue!15}{72.99} & \cellcolor{green!20}{\textbf{0.059}} & 0.004 & \cellcolor{blue!15}{\textbf{71.70}} & \cellcolor{green!20}{0.052} & \cellcolor{green!20}{0.003} & \cellcolor{blue!15}{55.21} & \cellcolor{green!20}{0.092} & \cellcolor{green!20}{0.010} & \cellcolor{blue!15}{53.87} \\
 & Full       & \cellcolor{green!20}{\textbf{0.022}} & \cellcolor{green!20}{\textbf{0.003}} & \cellcolor{blue!15}{\textbf{74.70}} & \cellcolor{green!20}{0.065} & 0.004 & \cellcolor{blue!15}{71.49} & \cellcolor{green!20}{\textbf{0.029}} & \cellcolor{green!20}{\textbf{0.002}} & \cellcolor{blue!15}{\textbf{57.18}} & \cellcolor{green!20}{\textbf{0.069}} & \cellcolor{green!20}{\textbf{0.004}} & \cellcolor{blue!15}{\textbf{53.99}} \\
\midrule
\multirow{3}{*}{Gemini-2.0-Flash}
 & Original   & \textbf{0.024} & 0.009 & 74.17 & \textbf{0.046} & 0.006 & 73.74 & 0.072 & \textbf{0.002} & 64.54 & \textbf{0.041} & \textbf{0.002} & 63.07 \\
 & Cyclic     & 0.035 & \cellcolor{green!20}{0.004} & \cellcolor{blue!15}{75.24} & 0.049 & \cellcolor{green!20}{\textbf{0.003}} & \cellcolor{blue!15}{\textbf{75.67}} & 0.077 & 0.004 & \cellcolor{blue!15}{66.62} & 0.058 & \cellcolor{green!20}{\textbf{0.002}} & \cellcolor{blue!15}{65.27} \\
 & Full       & 0.037 & \cellcolor{green!20}{\textbf{0.003}} & \cellcolor{blue!15}{\textbf{75.67}} & 0.061 & \cellcolor{green!20}{\textbf{0.003}} & \cellcolor{blue!15}{75.56} & \cellcolor{green!20}{\textbf{0.051}} & \cellcolor{green!20}{\textbf{0.002}} & \cellcolor{blue!15}{\textbf{67.97}} & 0.074 & 0.003 & \cellcolor{blue!15}{\textbf{65.52}} \\
\midrule
\multirow{3}{*}{Voxtral-Mini-3B}
 & Original   & 0.189 & 0.006 & 54.98 & \textbf{0.079} & 0.025 & 54.23 & 0.122 & 0.035 & 44.05 & 0.140 & 0.007 & 47.73 \\
 & Cyclic     & \cellcolor{green!20}{\textbf{0.091}} & \cellcolor{green!20}{0.005} & \cellcolor{blue!15}{59.81} & 0.113 & \cellcolor{green!20}{0.013} & \cellcolor{blue!15}{57.13} & \cellcolor{green!20}{\textbf{0.034}} & \cellcolor{green!20}{0.006} & \cellcolor{blue!15}{49.20} & \cellcolor{green!20}{\textbf{0.033}} & \cellcolor{green!20}{\textbf{0.001}} & \cellcolor{blue!15}{51.05} \\
 & Full       & \cellcolor{green!20}{0.118} & \cellcolor{green!20}{\textbf{0.002}} & \cellcolor{blue!15}{\textbf{63.35}} & 0.119 & \cellcolor{green!20}{\textbf{0.006}} & \cellcolor{blue!15}{\textbf{59.81}} & \cellcolor{green!20}{0.035} & \cellcolor{green!20}{\textbf{0.002}} & \cellcolor{blue!15}{\textbf{49.94}} & \cellcolor{green!20}{0.052} & \cellcolor{green!20}{0.002} & \cellcolor{blue!15}{\textbf{52.51}} \\
\midrule
\multirow{3}{*}{Voxtral-Small-24B}
 & Original   & 0.095 & 0.004 & 64.63 & \textbf{0.074} & 0.011 & 62.27 & 0.085 & 0.003 & 60.00 & 0.066 & \textbf{0.002} & 57.43 \\
 & Cyclic     & \cellcolor{green!20}{\textbf{0.088}} & \cellcolor{green!20}{0.002} & \cellcolor{blue!15}{66.78} & 0.123 & \cellcolor{green!20}{\textbf{0.005}} & \cellcolor{blue!15}{64.63} & \cellcolor{green!20}{0.085} & \cellcolor{green!20}{0.002} & \cellcolor{blue!15}{62.21} & \cellcolor{green!20}{\textbf{0.041}} & 0.003 & \cellcolor{blue!15}{58.65} \\
 & Full       & \cellcolor{green!20}{0.090} & \cellcolor{green!20}{\textbf{0.001}} & \cellcolor{blue!15}{\textbf{67.63}} & 0.107 & \cellcolor{green!20}{\textbf{0.005}} & \cellcolor{blue!15}{\textbf{66.46}} & \cellcolor{green!20}{\textbf{0.077}} & \cellcolor{green!20}{\textbf{0.001}} & \cellcolor{blue!15}{\textbf{62.33}} & \cellcolor{green!20}{\textbf{0.026}} & 0.003 & \cellcolor{blue!15}{\textbf{60.98}} \\
\bottomrule
\end{tabular}
\end{adjustbox}
\label{tab:permutation}
\vspace{-8pt}
\end{table*}

\subsection{Permutation to Alleviate Bias}
As shown in Figure~\ref{fig:fluctuation}, the performance of LALMs fluctuates considerably when the options are shuffled. 
Such sensitivity to ordering poses a critical threat to the reliability of benchmark results, since model accuracy may reflect structural bias rather than genuine reasoning ability. 
Hence, we apply permutation-based evaluation strategies, which have been widely adopted in prior work on LLMs \cite{izacard2023atlas, zheng2024large, liusie-etal-2024-teacher}, to investigate whether they can mitigate position bias and yield more reliable assessments of LALMs. 
Specifically, after shuffling the order of options, each permutation is treated as an independent input, and the final answer is determined through majority voting across all permutations. 
This approach closely resembles the idea of self-consistency~\cite{wangself}, except that the source of randomness comes from shuffling option order rather than sampling.

The results are presented in Table~\ref{tab:permutation}. 
Although RSD may fail to capture bias well when label distributions are unbalanced~\cite{choi-etal-2025-mitigating}, such as in MMAU and SPEECH-MMAU, we still report this metric for reference. 
Cyclic permutation yields higher performance than the original evaluation. 
In contrast, full permutation further improves on cyclic permutation and provides the most reliable estimate of a model’s true capability because it considers all possible answer orders. 
Even though Section~\ref{sec:lalms_llms} highlights that position bias may vary across text-only LLMs and audio-based LALMs, permutation still proves effective in mitigating this bias. 
It enables more reliable and trustworthy evaluation results for LALMs.

\begin{figure}[tbp]
    \centering
    \includegraphics[width=\linewidth]{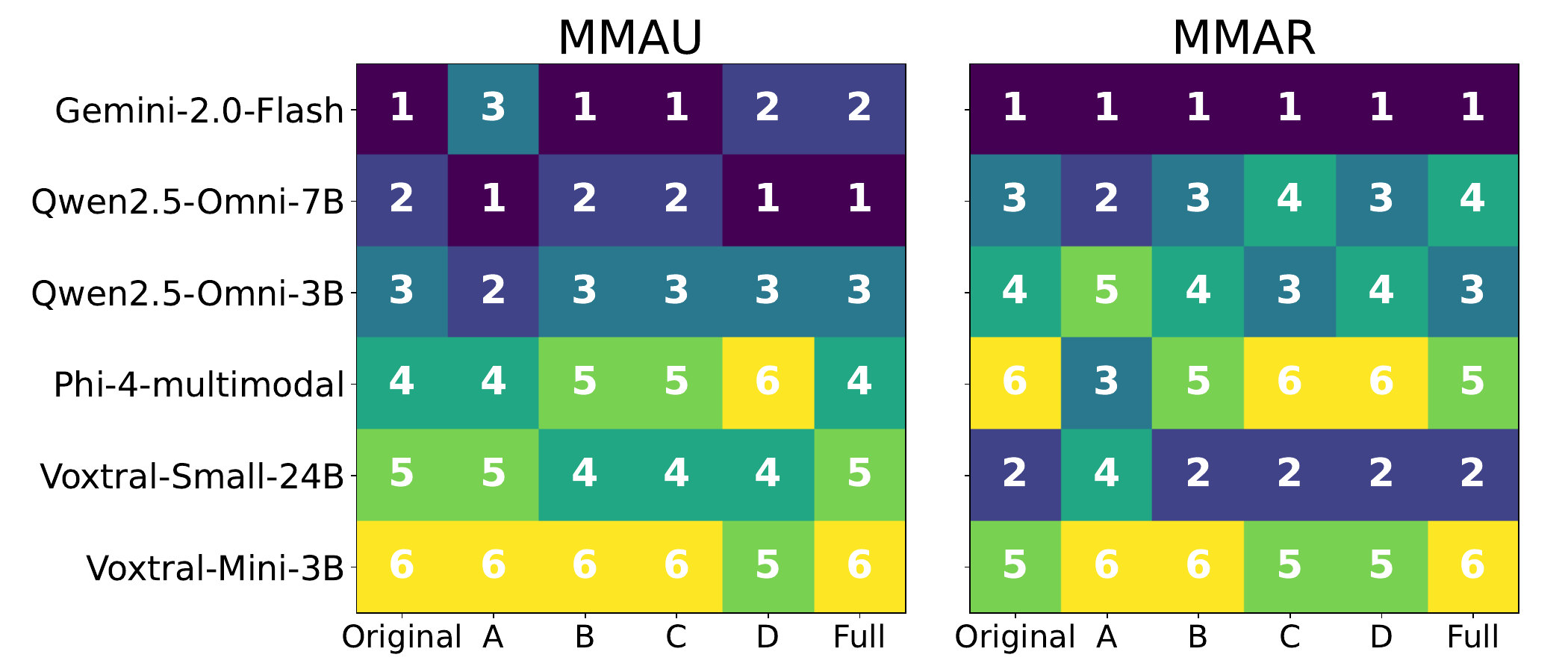}
    \caption{After shuffling the options, model rankings fluctuate considerably.}
    \vspace{-10pt}
    \label{fig:rank}
\end{figure}

\subsection{Model Ranking Fluctuation}
To further examine the impact of position bias, we compared the relative rankings of the six models after placing the gold answer in different positions, as shown in Figure~\ref{fig:rank}.

The results show that position bias can significantly alter model rankings. 
For example, in MMAU, Qwen2.5-Omni-7B may win over Gemini-2.0-Flash once their inherent bias are reduced. A similar pattern appears in MMAR, where Phi-4-Multimodal wins against Voxtral-Mini-3B, and Qwen2.5-Omni-3B also beats Qwen2.5-Omni-7B. 
These findings emphasize the importance of adopting full permutation in evaluation.

\section{Conclusion}
This work presents the first systematic investigation of position bias in LALMs. We show that these models are highly sensitive to option order, which can substantially distort their outputs and lead to unreliable behavior. While permutation-based methods help mitigate this issue, they incur additional computational overhead. Reducing position bias is therefore essential for improving the overall trustworthiness and reliability of LALMs. By identifying and characterizing this problem, our study establishes a foundation for future research to move beyond permutation, toward more advanced solutions that enable fairer and more efficient use of LALMs.

\section{Generative AI Use Disclosure}
We used large language models to polish the manuscript. The use of these AI tools did not influence the substantive content, data analysis, or scientific conclusions of the study and served exclusively as a writing aid. We have carefully checked the AI refinements to ensure that all results are correct.

\section{Acknowledgments}
We thank the reviewers for their constructive feedback during the review process, which greatly contributed to improving this work. We also acknowledge the computational and storage support provided by the National Center for High-performance Computing (NCHC) of the National Applied Research Laboratories (NARLabs) in Taiwan.
This work was supported by the Ministry of Education (MOE) of Taiwan under the project Taiwan Centers of Excellence in Artificial Intelligence, through the NTU Artificial Intelligence Center of Research Excellence (NTU AI-CoRE).

\bibliographystyle{IEEEtran}
\bibliography{mybib}

\end{document}